\documentclass{appolb}
\usepackage{epsfig}
\begin{document}
\title{Spectroscopy at HERA%
\thanks{Presented at PHOTON2005, 31.08-4.09.2005, Warsaw}%
}
\author{Kropivnitskaya Anna
\address{Institute for Theoretical and Experimental Physics}
}
\maketitle
\begin{abstract}

Recent results on spectroscopy with special focus on searches for
pentaquarks are presented from the H1 and ZEUS collaborations.
Cross sections
of observed states and upper limits on the production cross section of  
unobserved states are extracted in order to enable
comparison between experiments.
Measurements of the inclusive photoproduction of the neutral
mesons $\eta$, $\rho^0$, $f_0(980)$ and $f_2(1270)$ in $ep$ interactions at
HERA at an average $\gamma{p}$ collision energy  of $210$ GeV are also presented.

\end{abstract}
  
\section{Introduction}

Recently, 
some experiments 
have reported narrow signals in the vicinity of
 1530 MeV in the $nK^+$ and $pK^0_S$ invariant mass spectra which are consistent with
 the exotic pentaquark baryon state $\Theta^+$ with quark content $uudd\bar{s}$ \cite{3},
while other experiments have searched for this state with negative results. 
The possible existence a charm pentaquark has also been discussed,
with
renewed theoretical interest in calculating their expected properties~\cite{7,10} following
the observation of strange pentaquarks.

The results of pentaquark searches by the H1 and the ZEUS collaboration are presented
in this report. The analyses are based on data samples with an integrated luminosity of
75 pb$^{-1}$ and 126 pb$^{-1}$ taken by the H1 and the ZEUS collaborations in the years 
1994-2000 when HERA collided electrons or positrons with 27.6 GeV and protons with 820 or 920 GeV.

Understanding the process whereby quarks and gluons convert to colourless hadrons is
one of the outstanding problems in particle physics.
The measurements of the inclusive photoproduction of the neutral
mesons $\eta$, $\rho^0$, $f_0(980)$ and $f_2(1270)$ could contribute to resolving
this problem.

\section{Pentaquarks}

A strange pentaquark $\Theta^+$ candidate was seen by the ZEUS collaboration. 
In this analysis deep inelastic scattering ($DIS$) events were selected 
by requiring an exchanged photon virtuality $Q^2>1$ GeV$^2$.  
The $\Theta^+$ was reconstructed in the decay channel to $pK_S^0$. 
\begin{figure}[t]
\center
\setlength{\unitlength}{1cm}
\begin{picture}(6.0,5.7)
\epsfig{file=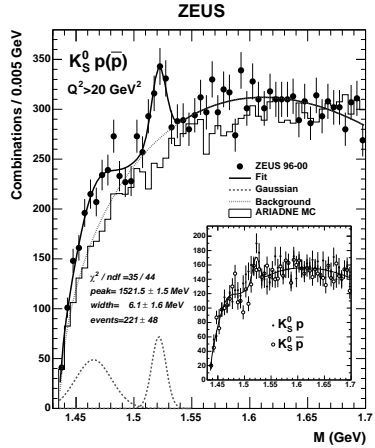,width=5.0cm}
\end{picture}
\caption{Invariant mass spectrum for the $K_S^0p(\bar p)$ channel
for $Q^2>20$ GeV$^2$. 
}
\label{theta}
\end{figure}
In Fig.~\ref{theta}, the invariant mass of proton and $K_S^0$ is presented
for events with $Q^2>20$ GeV$^2$. 
This distribution was fitted with a polynomial background and two Gaussians.
The signal peak is observed with a mass of $1521.5\pm 1.5^{+2.8}_{-1.7}$ 
MeV and the measured Gaussian width $6.1\pm1.6^{+2.0}_{-1.4}$ MeV which is
consistent with the detector resolution. The significance of the signal 
is 4.6 sigma. The second Gaussian significantly improves the fit in the low mass region,
and has a mass of about 1470 MeV and a width of 16 MeV and may correspond to the $\Sigma(1480)$
for which the evidence of existence is poor. 
The invariant-mass spectrum was investigated 
for the  $pK_S^0$ and $\bar{p}K_S^0$ separately. The signal has been seen in both charges
with significance of 3 $\sigma$. If the signal corresponds to the $\Theta^+$ pentaquark, 
this provides the first evidence for its antiparticle. 
The measured total cross section for the $\Theta^+$ in the kinematic region $Q^2>20$ GeV$^2$,
$p_T>0.5$ GeV, $|\eta|<1.5$ and $0.04<y_e<0.95$ is $125\pm 27^{+37}_{-28}$ pb.
The properties of the $\Theta^+$ candidate were studied and 
it was found to be produced predominantly 
in the forward pseudorapidity region.

A similar analysis was done by the H1 collaboration and no peak is visible near 1520 MeV. 
The resulting upper limit on the $\Theta^+$ production cross section was found to vary between 
40 and 120 pb  over the mass range of 1.48 to 1.7 GeV 
and
does not exclude the previously measured cross section at ZEUS.

\begin{figure}[t]
\center
\setlength{\unitlength}{1cm}
\begin{picture}(10.0,5.8)
\epsfig{file=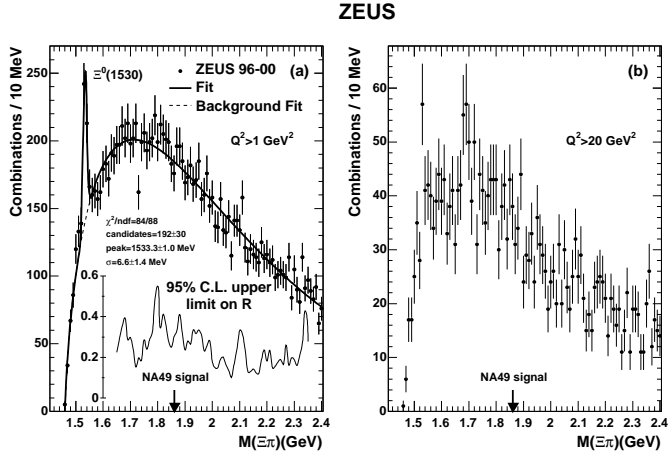,width=9.0cm}
\end{picture}
\caption{The $\Xi\pi$ invariant-mass spectrum for: (a) $Q^2>1$ GeV$^2$ and,
(b) $Q^2>20$ GeV$^2$ (all four charge combinations summed).
}
\label{Xi}
\end{figure}

NA49 collaboration published 
results of double strange pentaquark $\Xi^{--}_{3/2}$ 
searches in the $\Xi\pi$ invariant mass spectrum~\cite{6}. 
They have observed a narrow peak with mass of about 1860 MeV.
A similar analysis was repeated using ZEUS
 $DIS$ data. The $\Xi^{--}_{3/2}$ was reconstructed from $\Xi^-\pi^-$ decay. 
ZEUS observed a clean signal of $\Xi^0(1530)$ 
in $DIS$ events with $Q^2>1$ GeV$^2$, but no signal with a mass around
1860 MeV is observed (Fig.~\ref{Xi}). A similar analysis was performed for $Q^2>20$ 
GeV$^2$, the kinematic region where the $\Theta^+$ state was 
most clearly observed by ZEUS. Again no signal is observed near 1860 MeV. 
The number of $\Xi^0(1530)$ signal events reconstructed in this analysis 
is about the same as for NA49 data. However, it should be noted that NA49 
is a fixed target experiment, which has good acceptance in the forward 
region. The non-observation of this signal in the central-fragmentation 
region in the ZEUS data does not necessarily contradict the observation 
of a signal predominantly produced in the forward region. 

\begin{figure}[t]
\center
\setlength{\unitlength}{1cm}
\begin{picture}(6.0,3.5)
\epsfig{file=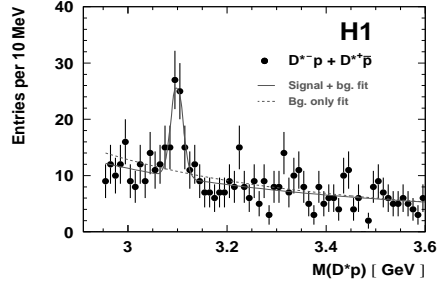,
        bburx=530,
        bbury=410,bbllx=50,bblly=60,clip=,
        height=3.7cm,width=5.5cm,angle=0}
\end{picture}
\caption{$M(D^*p)$ distribution from opposite-charge $D^*p$ combinations in $DIS$.
}
\label{Thetac}
\end{figure}

A search for a charm pentaquark $\Theta_c^0$ candidate was carried out 
by H1 using $DIS$ data at $Q^2>1$ GeV$^2$. 
The $\Theta_c^0$ was reconstructed via its decay to $D^*p$,
where the $D^*$ is reconstructed using the decay channel $D^*\to D^0\pi \to K\pi\pi$. 
A clear and narrow resonance is observed for both $D^{*-}p$ and 
$D^{*+}\bar{p}$ combinations with an invariant mass of 
$M(D^*p) = 3099\pm3\pm5$ MeV (Fig.~\ref{Thetac}) 
with a significance of 5.4$\sigma$.
The measured width of the resonance is $12\pm3$(stat.) MeV, 
consistent with the experimental resolution. 
A signal with compatible
mass and width is also observed in an independent photoproduction data 
sample.
An acceptance corrected ratio  
$R_{corr}(D^*p(3100)/D^*)=1.59\pm0.33^{+0.33}_{-0.45}$ $\%$ was measured. 
When extrapolating to the full phase space of the decay products one observes 
a ratio of the visible cross section of 
$\sigma_{vis}(D^*p(3100))/\sigma_{vis}(D^*)=(2.48\pm0.52^{+0.85}_{-0.64})$ $\%$.
The region of $M (D^*p)$ in which the signal is 
observed contains a richer yield of $D^*$ mesons and exhibits a harder
proton candidate momentum distribution than is the case for side bands in
$M(D^*p)$. 
Compared to inclusive $D^*$ production the $D^*p(3100)$ production seems 
to be suppressed in the close to central rapidity regions in laboratory and 
centre-of-mass frames.

A similar analysis was done by ZEUS using higher statistics
and reconstructing $D^*$ mesons in two channels $D^*\to D^0\pi \to K\pi\pi$ 
and $D^*\to D^0\pi \to K\pi\pi\pi\pi$.
No signal near 3100 MeV is observed. 
ZEUS estimated the upper limit on the acceptance corrected ratio and it is equal 0.59 $\%$ 
(0.51 $\%$ for both $D^*$ decay channels) in $DIS$.

\section{Light Mesons ($\eta$, $\rho^0$, $f_0(980)$, $f_2(1270)$) Photoproduction}

Besides exotic searches for particles the 
production of  well-known hadrons such as pions, $K^0_S$, $\Lambda$, protons, 
charm mesons, $J/\psi$, {\it etc.} are measured by ZEUS and H1. 
A recent result is the cross section
measurement of  inclusive photoproduction of $\eta$, $\rho^0$, $f_0(980)$ 
and $f_2(1270)$ mesons at H1 in the central rapidity region. 
In this analysis a photoproduction data sample taken in the year
2000 corresponding to an integrated luminosity of 38.7 pb$^{-1}$ 
and average $\sqrt{S_{\gamma p}}=$210 GeV was used. 
The $\rho^0$, $f_0(980)$ and $f_2(1270)$ mesons were 
reconstructed though $\pi^+\pi^-$ decay, $\eta$ meson 
though $\gamma\gamma$ decay.

\begin{figure}[t]
\center
\setlength{\unitlength}{1cm}
\begin{picture}(6.0,5.7)
\epsfig{file=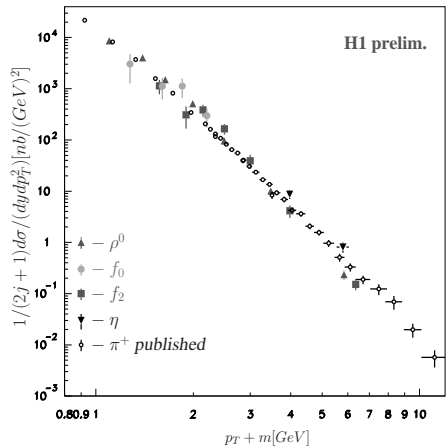,
        bburx=530,
        bbury=630,bbllx=100,bblly=210,clip=,
        height=6.0cm,width=6.0cm,angle=0}
\end{picture}
\caption{The differential photoproduction cross section of 
$\eta$, $\rho^0$, $f_0(980)$ and $f_2(1270)$. 
}
\label{sigma}
\end{figure}

In Fig.~\ref{sigma}, the measured differential cross section 
$1/(2j+1) d^2\sigma/dydp_T^2$ of the resonances as a function of $m+p_T$,
where $j$ is a spin and $m$ is a mass of measured particle,
was compared with the cross 
section of charged pions. The resonances have a similar 
behavior as observed for long-lived hadrons~\cite{rostov}. 

\section{Conclusions}

The ZEUS collaboration observes a narrow baryonic state at a mass of about 1520 MeV
that is interpreted as strange pentaquark $\Theta^+$.
H1 does not observe this state but the upper limit
on the production cross section does not 
exclude the ZEUS observation.
A resonance search has been performed in the $D^*p$ invariant-mass
spectrum with the H1 and the ZEUS detector. H1 observes a $\Theta^0_c$
candidate at mass of about 3100 MeV. ZEUS does not see any signal 
near this mass value. The upper limit on the $\Theta^0_c$ production
contradicts the H1 observation.
More data from HERA{\it II} which are now being taken should 
resolve this contradiction.

The inclusive cross section for $\eta$, $\rho^0$, $f_0(980)$ and
$f_2(1270)$ was measured and 
has the similar behavior as observed for long-lived hadrons~\cite{rostov}.

\section{Acknowledgments}
This work was partially supported by grants RFBR 03-02-17291, 04-02-16445 and CRDF REC-011.
I would like to thank Dmitry Ozerov and Andrey Rostovtsev for useful discussion.


\begin{thebibliography}{99}

\bibitem{3}
    D. Diakonov, V. Petrov, M. V. Polyakov, Z. Phys. {\bf A359}, 305 (1997).

\bibitem{7}
 R. L. Jaffe
and F. Wilczek, Phys. Rev. Lett. {\bf 91}, 232003 (2003). 

\bibitem{10}
M. Karliner and H. J. Lipkin, [hep-ph/0307343];\\
K. Cheung, [hep-ph/0308176].

\bibitem{6}
 C. Alt et al. [NA49 Collaboration], Phys. Rev. Lett. {\bf 92}, 042003 (2004).

\bibitem{rostov}
A.~Rostovtsev, In Proceedings of the 31st International
Symposium on Multiparticle Dynamics (ISMD
2001), Datong, China, 1-7 Sep 2001.

\end{thebibliography}
\end{document}